\documentclass[a4paper,fleqn,usenatbib,useAMS]{mnras}

\usepackage{newtxtext,newtxmath}
\usepackage[T1]{fontenc}
\usepackage{ae,aecompl}

\usepackage{graphicx}	% Including figure files
\usepackage{amsmath}	% Advanced maths commands
\usepackage{tikz}       % to define some graphic symbols

\newcommand\PI{3.14159} % Difine PI constant for calculation purposes

\newcommand{\circlesym}{%
	\begin{tikzpicture}[x=0.8ex,y=0.8ex,baseline={([yshift=-0.8ex]current bounding box.center)}]
	\draw (0,0) circle [radius=1] ;
	\end{tikzpicture}}

\newcommand{\squaresym}{%
	\begin{tikzpicture}[x=0.8ex,y=0.8ex,baseline={([yshift=-0.8ex]current bounding box.center)}]
	\draw (0,0) rectangle ({sqrt(\PI)},{sqrt(\PI)}) ;
	\end{tikzpicture}}

\newcommand{\diamondsym}{%
	\begin{tikzpicture}[x=0.8ex,y=0.8ex,baseline={([yshift=-0.8ex]current bounding box.center)}]
	\draw[rotate=45] (0,0) rectangle ({sqrt(\PI)},{sqrt(\PI)}) ;
	\end{tikzpicture}}

\newcommand{\trianglupsym}{%
	\begin{tikzpicture}[x=0.8ex,y=0.8ex,baseline={([yshift=-0.8ex]current bounding box.center)}]
	\draw (-1,-1) -- (0,1) -- (1,-1) -- cycle ;
	\end{tikzpicture}}

\newcommand{\triangldnsym}{%
	\begin{tikzpicture}[x=0.8ex,y=0.8ex,baseline={([yshift=-0.8ex]current bounding box.center)}]
	\draw (-1,1) -- (1,1) -- (0,-1) -- cycle ;
	\end{tikzpicture}}

\newcommand{\starsym}{%
	\begin{tikzpicture}[x=0.9ex,y=0.9ex,baseline={([yshift=-0.8ex]current bounding box.center)}]
	\draw (45:0.4)
	\foreach \i in {1,...,4}{ -- (\i*90:1) -- (\i*90+45:0.4)}
	-- cycle;
	\end{tikzpicture}}

\title[LBV candidates in M31]{Luminous Blue Variable candidates in M31}

\author[A. Sarkisyan et al.]{
A. Sarkisyan,$^{1}$\thanks{E-mail: ars@sao.ru}
O. Sholukhova,$^{1}$
S. Fabrika,$^{1,4}$
D. Bizyaev,$^{2,3}$
A. Valeev,$^{1,4}$
\newauthor
A. Vinokurov,$^{1}$
Y. Solovyeva,$^{1}$
A. Kostenkov,$^{1,5}$
V. Malanushenko$^{2}$
and P. Nedialkov$^{6}$
\\
$^{1}$Special Astrophysical Observatory of the Russian Academy of Sciences (SAO RAS), Nizhnij Arkhyz, Karachai-Cherkessia, Russia\\
$^{2}$Apache Point Observatory and New Mexico State University, Sunspot, NM, 88349, USA\\
$^{3}$Sternberg Astronomical Institute, Moscow State University, Moscow, Russia\\
$^{4}$Kazan (Volga region) Federal University, Kazan, Russia\\ 
$^{5}$Saint Petersburg State University, Saint Petersburg, Russia\\ 
$^{6}$Department of astronomy, Sofia University, 5 J. Bourchier blvd, Sofia 1164, Bulgaria
}

\date{Accepted XXX. Received YYY; in original form ZZZ}

\pubyear{2020}

\begin{document}
\label{firstpage}
\pagerange{\pageref{firstpage}--\pageref{lastpage}}
\maketitle

\begin{abstract}

We study five Luminous Blue Variable (LBV) candidates in the Andromeda galaxy and one more (MN112) in the Milky Way. We obtain the same-epoch near-infrared (NIR) and optical spectra on the 3.5-meter telescope at the Apache Point Observatory and on the 6-meter telescope of the SAO RAS. The candidates show typical LBV features in their spectra: broad and strong hydrogen lines, \ion{He}{i}, \ion{Fe}{ii}, and [\ion{Fe}{ii}] lines. We estimate the temperatures, reddening, radii and luminosities of the stars using their spectral energy distributions. Bolometric luminosities of the candidates are similar to those of known LBV stars in the Andromeda galaxy. One candidate, J004341.84+411112.0, demonstrates photometric variability (about 0.27\,mag in \textit{V} band), which allows us to classify it as a LBV. The star J004415.04+420156.2 shows characteristics typical for B[e]-supergiants. The star J004411.36+413257.2 is classified as Fe\,II star. We confirm that the stars J004621.08+421308.2 and J004507.65+413740.8 are warm hypergiants. We for the first time obtain NIR spectrum of the Galactic LBV candidate MN112. We use both optical and NIR spectra of MN112 for comparison with similar stars in M31 and notice identical spectra and the same temperature in the J004341.84+411112.0. This allows us to confirm that MN112 is a LBV, which should show its brightness variability in longer time span observations. 

\end{abstract}


%%%%%%%%%%%%%%%%%%%%%%%%%%%%%%%%%%%%%%%%%%%%%%%%%%

%%%%%%%%%%%%%%%%% BODY OF PAPER %%%%%%%%%%%%%%%%%%

\section{Introduction}

Classification of highest luminosity stars from the most upper part of the HR diagram demonstrates a great improvement during last years, partly due to works by R. Humphreys and colleagues \citep{Humphreys2013,Humphreys2014,Humphreys2017a,Humphreys2017b,Gordon2016} 
who studied such stars in nearby galaxies M31 and M33. Reliable classification of the highest luminosity stars requires long-term photometric and spectral monitoring in the optical and near-infrared (O/NIR) parts of spectra. 
We have presented a new verification method and investigation of 6 LBV candidates in M31 in the O/NIR \citep{Sholukhova2015}, and here we continue with investigating five more luminous stars in the Andromeda, among with one more in the Milky Way galaxy.  

\cite{Humphreys2014} split the highest luminosity stars into six types using their spectral and photometric features: LBV stars and LBV candidates, B[e]-supergiants (B[e]SGs), Fe\,II emission-line stars, warm hypergiants, 
hot and intermediate-type (yellow) supergiants and Of/late-WN stars. A brief outline of their features follows.

\paragraph*{LBV stars and LBV candidates}
  
They are evolved and unstable high luminosity stars. In the maximum of their brightness LBVs are A-F hypergiants, while in the minimum they can show WNLh spectra. In an intermediate state they mimic B[e] supergiants. There is a variety of observational manifestations of LBVs because of their strong stellar wind. This evolutionary stage is characterized by the depletion of hydrogen in nuclei. Changes in the ionization state of the most abundant elements in the stars lead to the variations of the gas transparency and stellar mass loss rate. The most distinctive feature that points at the true LBV nature is their S Dor-type variability, when the stars show aperiodic brightness variation that exceeds 0.2\,mag \citep{vanGenderen2001} on the time scales from months to decade, accompanied by a color variability. Although there is no common paradigm for the explanation of those features, 
they should be primarily caused by variations of stellar radius and temperature, and not by changes of the dust extinction. LBVs do not have [\ion{O}{i}] 6300, 6364\,\AA\ lines in their spectra. Some of them show emission lines \ion{Fe}{ii} and [\ion{Fe}{ii}]. The spectra show free-free emission, but not traces of warm emission or infrared (IR) excess. 
\cite{Humphreys2014} showed that confirmed LBVs have relatively low speed of stellar wind in their hot state 
(which corresponds to the visual brightness minimum) with respect to the Of/WN stars, which they may resemble. 

\paragraph*{B[e]-supergiants}

A connection between the LBVs and B[e]SGs is still unclear \citep{Kraus2014,Kraus2019}, although these classes of massive stars have similar luminosity and spectra when the LBVs are in their hot phase. At the same time, physical properties of LBVs and B[e]SGs can be very different \citep{Fabrika2000}. LBVs can increase their photosphere size ten times and brightness by 3 magnitudes, while the B[e]SGs do not show strong brightness variability. 
The main spectral difference from the LBVs is the presence of the [\ion{O}{i}] 6300, 6364\,\AA\ and [\ion{Fe}{ii}] emission lines in their spectra. In addition to numerous [\ion{Fe}{ii}] emission and broad hydrogen emission lines they also show [\ion{Ca}{ii}] 7291, 7324\,\AA\ emission lines. Recently \cite{Aret2016} pointed at the [\ion{O}{i}] 6300, 6364\,\AA\ emission lines as one of the main features of the B[e]SG class, while the [Ca II] 7291, 7324\,\AA\ emission being dependent of the presence of circumstellar gas is detected in the more massive stars often, but is not necessarily seen in all B[e]SGs. Confirmed B[e]SG stars have warm circumstellar dust, which their spectral energy distributions (SED) reveal: most of the B[e]SGs demonstrate the infrared excess over the amount expected from the free-free radiation of their stellar winds. The latter is also the main criterion that allows us to distinguish the B[e]SGs from LBVs \citep{Oksala2013,Kraus2014,Humphreys2017a}.  

\paragraph*{Fe\,II emission line stars}

These stars have blue continuum with strong hydrogen emission lines, and \ion{Fe}{ii} emission lines. At the same time, they do not show any absorption lines, [\ion{O}{i}] emission or circumstellar dust. Their nature is not yet clear. Similar to the B[e]SGs, these stars do not show significant variability of spectra. 

\paragraph*{Warm Hypergiants}

Spectra of the warm hypergiants are similar to those of LBVs at their bright state. They are more luminous than the supergiants. They show strong Balmer emission with broad wings and P Cygni profile, in combination with Ca\,II (8498, 8542, 8662\,\AA) and [\ion{Ca}{ii}] 7291, 7324\,\AA\ emission lines. All these stars are surrounded by dust and show strong infrared excess. Some of them indicate the [\ion{O}{i}] and [\ion{Fe}{ii}] emission similar to the B[e]SGs. They are different from the supergiants by the presence of the A- and F-type absorption features. The hypergiants do not show spectral variability. 

\paragraph*{Hot Supergiants and Yellow Supergiants}

Their spectra show mostly the hydrogen emission lines, and some have \ion{He}{i} emission. Many lines have P Cygni profiles. \cite{Humphreys2014} notice that the outflow velocity in confirmed LBVs is significantly less than that in the hot supergiants. Intermediate class stars, or yellow supergiants, do not show spectral variability or \ion{Fe}{ii} and [\ion{Fe}{ii}] emission. The same is true for the supergiants of later types with A-spectra and yellow supergiants.  

\paragraph*{Of/late-WN stars} 

Some LBV stars like He 3-591 in our Galaxy, HD 5980 in SMC, R127, R71, HDE 269582 in LMC, AF And, Var 15 in M31, Var B, Var 2 in M33 show the spectra and optical variability similar with Of/late-WN stars. However it does not automatically mean that all Ofpe/WN stars are LBV candidates. Only specific S Dor-type variability allows ones to classify those stars as LBVs (e.g. as in the case of Romano's star (V532) in M33, \citet{Romano1978}). Short time changes in spectra also indicate that an Ofpe/WN star can be a LBV candidate. \cite{Humphreys2014} show that Of/WN stars have high stellar wind speed (over 300\,km\,s$^{-1}$) with respect to LBVs.

LBV is a narrow class of stars. Their study in our galaxy is difficult because of large dust extinction in the galactic plane and high uncertainty in the distance determination. This makes nearby galaxies especially valuable for the studies of LBVs.

Spectra of LBVs and related stars are sufficiently well researched in the optical region (see e.g. \citet{Walborn2000}). But IR spectra of LBVs are understudied. Just a few IR spectra of galactic LBVs were investigated by \citep{Hamann1994,Morris1996,Voors2000,Groh2007}. \cite{Kraus2014,Kourniotis2018} published IR spectra for 8 LBV candidates in M31 and M33. They classified two those candidates as B[e]SGs for the presence of $^{12}$CO lines. 

We presented optical and NIRs spectra, SEDs, and determined temperatures, reddening, stellar radii and luminosity of five LBV candidates and two known LBVs (AE And and Var A-1) in M31 \citep{Sholukhova2015}. Two of those LBV candidates, J004526.62\penalty0+415006.3 and J004051.59\penalty0+403303.0, were justified as LBVs. Two more candidates (J004417.10\penalty0+411928.0 and J004444.52\penalty0+412804.0) turned out to be B[e]SGs. The star J004350.50\penalty0+414611.4 remains a LBV candidate. 

In the present paper we report results of O/NIR spectral and photometric observations of five LBV candidates in the Andromeda galaxy, and of one more object of this kind in our Galaxy, MN112 \citep{Gvaramadze2010}. We select objects J004341.84\penalty0+411112.0, J004411.36\penalty0+413257.2, J004415.00\penalty0 +420156.2, J004507.65\penalty0+413740.8 and J004621.08\penalty0+421308.2 from the list of LBV candidates by \cite{Massey2007,Massey2016} for this study. We refer to the stars by their first (RA) coordinate throughout the paper. Below we describe their spectra, SEDs, and use them to determine the stellar parameters for the consequent classification.  

\section{Observations}

The optical spectra of all objects were obtained with the SCORPIO spectrograph \citep{AfanMois2005} on the 6-m telescope BTA SAO RAS in October-November 2012. The observing log is shown in \autoref{tab:spec}. The spectra reduction and extraction were performed with the \textsc{spextra} package designed to work with stars in crowded fields by \citet{Sarkisyan2017}. Several sets of \textit{BVRI} or \textit{BVR} photometry was also obtained with the SCORPIO for each object (commonly simultaneously with the spectra; see \autoref{tab:phot} for details).

\begin{table*}
	\begin{center}
		\caption{The spectroscopic observations. 
		The columns represent the date of the observations, average seeing, used instrument (or instrument and grism combination) for spectroscopic observations. Abbreviation SCO is used for SCORPIO spectrograph.
		Spectral ranges and resolution as full width at half maximum (FWHM) of the used instruments are as follows: 
		TripleSpec -- 0.95--2.46\,\micron, 5\,\AA;
		SCO/VPHG550G -- 3500--7200\,\AA, 11\,\AA;
		SCO/VPHG1200G -- 4000--5700\,\AA, 5.3\,\AA;
		SCO/VPHG1200R -- 5700--7500\,\AA, 5.3\,\AA.}\label{tab:spec}		
		\begin{tabular}{cccc}
			\hline\hline
			Object & Date & Seeing (arcsec) & Instrument/Grism \\
			\hline
            J004341.84 & 17.10.12 & 1.0 & TripleSpec    \\
                       & 15.10.12 & 1.4 & SCO/VPHG550G  \\
                       & 16.10.12 & 1.1 & SCO/VPHG1200G \\
            J004411.36 & 07.11.12 & 1.1 & TripleSpec    \\
                       & 17.10.12 & 1.2 & SCO/VPHG550G  \\
                       & 21.10.12 & 1.1 & SCO/VPHG1200G \\
            J004415.00 & 13.11.12 & 1.1 & TripleSpec    \\
                       & 18.10.12 & 2.4 & SCO/VPHG550G  \\
                       & 04.09.15 & 1.3 & SCO/VPHG550G  \\
                       & 22.10.12 & 1.3 & SCO/VPHG1200G \\
            J004507.65 & 17.10.12 & 1.0 & TripleSpec    \\
                       & 14.10.12 & 2.0 & SCO/VPHG550G  \\
                       & 17.10.12 & 1.3 & SCO/VPHG1200G \\
            J004621.08 & 13.11.12 & 1.2 & TripleSpec    \\
                       & 20.10.12 & 0.9 & SCO/VPHG550G  \\
                       & 04.09.15 & 1.3 & SCO/VPHG1200G \\
            MN112      & 13.11.12 & 1.2 & TripleSpec    \\
                       & 16.08.15 & 4.0 & SCO/VPHG550G  \\
                       & 20.06.09 & 1.4 & SCO/VPHG1200G \\
                       & 20.06.09 & 1.4 & SCO/VPHG1200R \\
			\hline
		\end{tabular}
	\end{center}
\end{table*}

The NIR spectra are obtained with the TripleSpec \citep{Wilson2004} spectrograph on the 3.5-m ARC telescope at the Apache Point Observatory (APO, New Mexico, USA) in October and November of 2012 (\autoref{tab:spec}). The data reduction was performed with the \textsc{spextool} \cite{Cushing2004} package coupled with the \textsc{xtellcor} method developed by \cite{Vacca2003}.

We also were able to obtain quasi-simultaneous NIR and optical photometry for our LBV candidates in October-November of 2016. Each of the studied candidates was observed during one clear photometric one night on October 26, 2016, with the NICFPS imager on the 3.5-m ARC telescope at the APO. The imager is equipped with the MKO \textit{J}, \textit{H} \& \textit{K$_s$} filter set \citep{SimonsTokunaga2002,Tokunaga2002,TokunagaVacca2005}. 
Typically, we performed 3x60\,sec exposures in the \textit{J} filter, 6x20\,sec exposures in the \textit{H}, and 6x20\,sec in the \textit{K$_s$}. We applied spatial dithering between the exposures. All dark subtracted images were used to construct the flat field images by median co-adding them in the \textsc{iraf}, in each filter separately. After the flat fielding, magnitudes of observed stars were estimated by comparison with seven stars near the candidates in the NIR frames, with the help of Two Micron All Sky Survey (2MASS) calibration.

We were able to perform the optical photometry two weeks after the NIR observations with a small, 0.5-m ARCSAT telescope at APO, during clear, photometric nights on November 8 \& 14, 2016. We observed the stellar fields with a wide-field (32\,arcmin) ARCSAT camera equipped with the standard Johnson-Cousins \textit{BVRI} set of filters. 
Every night we took the Landolt fields of standards SA92 and PH0220 \citep{Landolt1992} for calibration purposes. 
The typical exposure time was 3x300\,sec in the \textit{B} band and 3x240\,sec in the \textit{V}, \textit{R} and \textit{I}. We applied spatial dithering between the exposures. We obtained sky flat field exposures, dark current frames and biases for each observing night. The images were bias and dark subtracted and flat fielded using standard \textsc{iraf} routines. We summarise results of our O/NIR photometry in \autoref{tab:phot}.

We also used here one additional spectrum of J004415.00 for its SED modeling. It was obtained as a byproduct of QSOs search on 3.5-m the ARC telescope at APO with the Dual Imaging Spectrograph (DIS) on 2018 October 7 using the 300\,line\,mm$^{-1}$ grating, which provided wavelength coverage from 5190 to 9850\,\AA\ and a resolution of roughly 5\,\AA.

\begin{table*}
	\caption{The O/NIR photometry. We show the date of the observations, used instrument (SCO for the SCORPIO/6-m BTA, and APO for the NICFPS/3.5m APO) and estimated magnitudes with their uncertainties.}\label{tab:phot}
	
	\begin{tabular}{c*{9}{|c}}
		\hline\hline 
		Object & Date & Instr. & \textit{B} & \textit{V} & \textit{R} & \textit{I} & \textit{J} & \textit{H} & \textit{K$_s$} \\ 
		\hline 
J004341.84 & 16.10.2012 & SCO & 18.02$\pm$0.05 & 17.59$\pm$0.03 & 17.12$\pm$0.05 &                &                &                &                \\ 
           & 17.01.2015 & SCO & 17.98$\pm$0.08 & 17.48$\pm$0.08 & 17.02$\pm$0.05 & 16.81$\pm$0.05 &                &                &                \\
           & 26.09.2016 & SCO & 17.83$\pm$0.08 & 17.32$\pm$0.05 & 16.91$\pm$0.06 &                &                &                &                \\
           & 26.10.2016 & APO &                &                &                &                &                & 15.85$\pm$0.24 & 15.73$\pm$0.06 \\
J004411.36 & 17.10.2012 & SCO & 18.61$\pm$0.07 & 17.89$\pm$0.05 & 17.38$\pm$0.07 &                &                &                &                \\ 
           & 17.01.2015 & SCO & 18.74$\pm$0.03 & 18.02$\pm$0.04 & 17.44$\pm$0.03 & 16.95$\pm$0.08 &                &                &                \\
           & 26.10.2016 & APO &                &                &                &                & 16.24$\pm$0.11 & 15.71$\pm$0.04 & 15.40$\pm$0.06 \\
J004415.00 & 18.10.2012 & SCO & 18.56$\pm$0.04 & 18.28$\pm$0.04 & 17.24$\pm$0.03 &                &                &                &                \\ 
           & 17.01.2015 & SCO & 18.57$\pm$0.06 & 18.26$\pm$0.05 & 17.24$\pm$0.06 & 17.23$\pm$0.06 &                &                &                \\
           & 04.09.2015 & SCO & 18.57$\pm$0.05 & 18.24$\pm$0.02 & 17.20$\pm$0.04 & 17.22$\pm$0.03 &                &                &                \\
           & 08.11.2016 & APO & 18.53$\pm$0.07 & 18.25$\pm$0.07 &                &                &                &                &                \\
           & 14.11.2016 & APO &                &                & 17.21$\pm$0.08 & 17.25$\pm$0.10 &                &                &                \\
           & 26.10.2016 & APO &                &                &                &                & 16.63$\pm$0.05 & 15.86$\pm$0.10 & 14.52$\pm$0.03 \\
J004507.65 & 14.10.2012 & SCO & 16.50$\pm$0.11 & 16.09$\pm$0.04 & 16.04$\pm$0.05 &                &                &                &                \\ 
           & 17.01.2015 & SCO & 16.52$\pm$0.06 & 16.18$\pm$0.03 & 15.96$\pm$0.03 & 15.73$\pm$0.10 &                &                &                \\
           & 08.11.2016 & APO & 16.48$\pm$0.04 & 16.20$\pm$0.07 & 16.00$\pm$0.04 &                &                &                &                \\
           & 14.11.2016 & APO &                &                &                & 15.89$\pm$0.11 &                &                &                \\
J004621.08 & 21.10.2012 & SCO & 18.40$\pm$0.06 & 18.05$\pm$0.01 & 17.65$\pm$0.04 &                &                &                &                \\ 
           & 17.01.2015 & SCO & 18.26$\pm$0.07 & 17.88$\pm$0.06 & 17.54$\pm$0.06 & 17.34$\pm$0.07 &                &                &                \\
           & 17.08.2015 & SCO &                & 17.95$\pm$0.06 & 17.62$\pm$0.06 & 17.34$\pm$0.06 &                &                &                \\
           & 04.09.2015 & SCO & 18.31$\pm$0.03 & 17.94$\pm$0.05 & 17.58$\pm$0.03 & 17.30$\pm$0.03 &                &                &                \\
           & 08.11.2016 & APO & 18.30$\pm$0.05 & 18.04$\pm$0.05 & 17.58$\pm$0.11 & 17.36$\pm$0.06 &                &                &                \\
           & 26.10.2016 & APO &                &                &                &                &                & 16.22$\pm$0.12 & 15.21$\pm$0.09 \\
MN112      & 20.06.2009 & SCO & 16.83$\pm$0.05 & 14.43$\pm$0.05 & 12.63$\pm$0.05 &                &                &                &                \\ 
           & 17.08.2015 & SCO & 16.84$\pm$0.06 & 14.43$\pm$0.08 & 12.65$\pm$0.07 &                &                &                &                \\
		\hline 
	\end{tabular} 
\end{table*}

\section{Results}

\subsection{Spectroscopy}

\begin{figure*}
	\includegraphics[width=\textwidth]{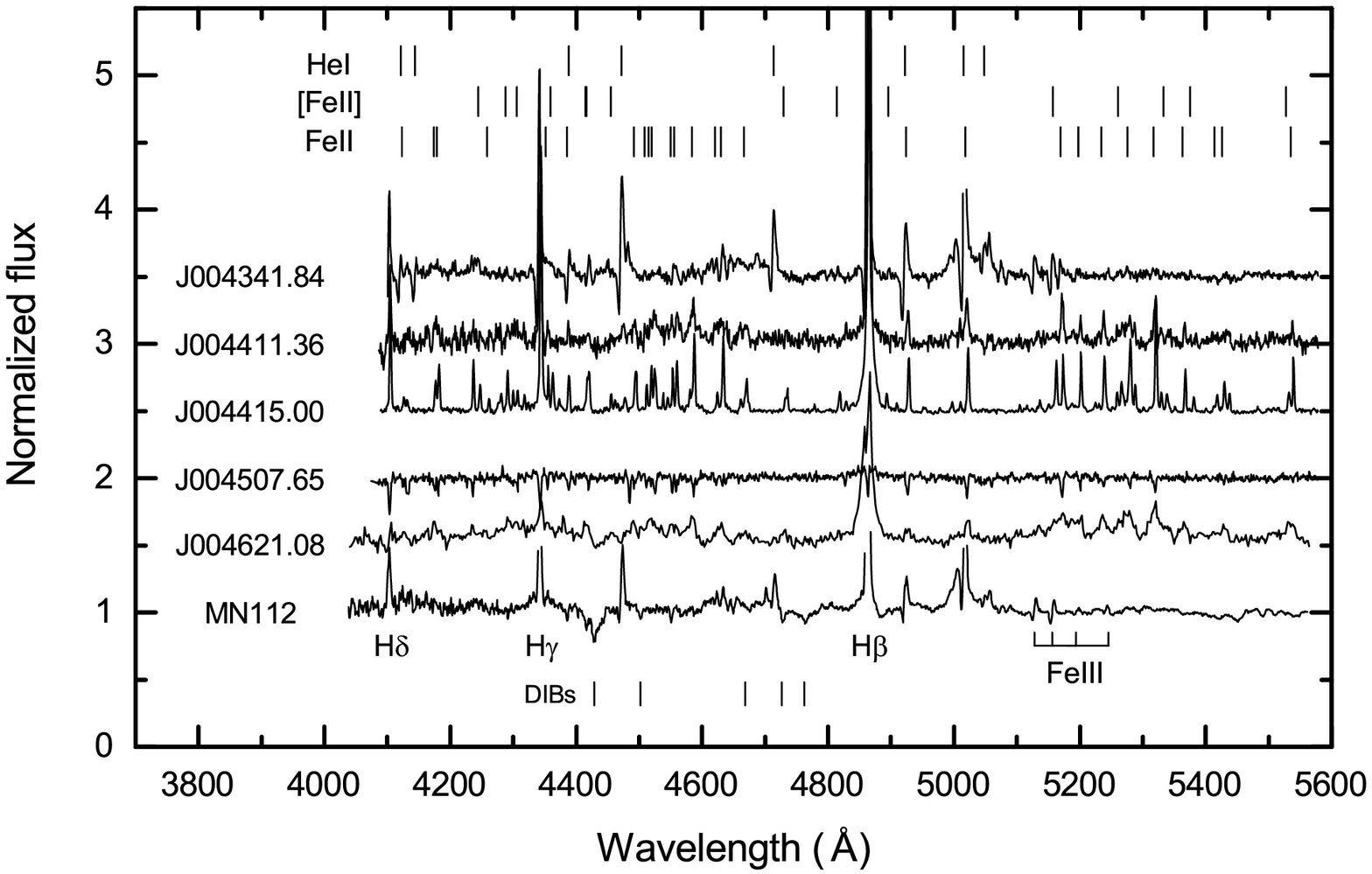}
	\includegraphics[width=\textwidth]{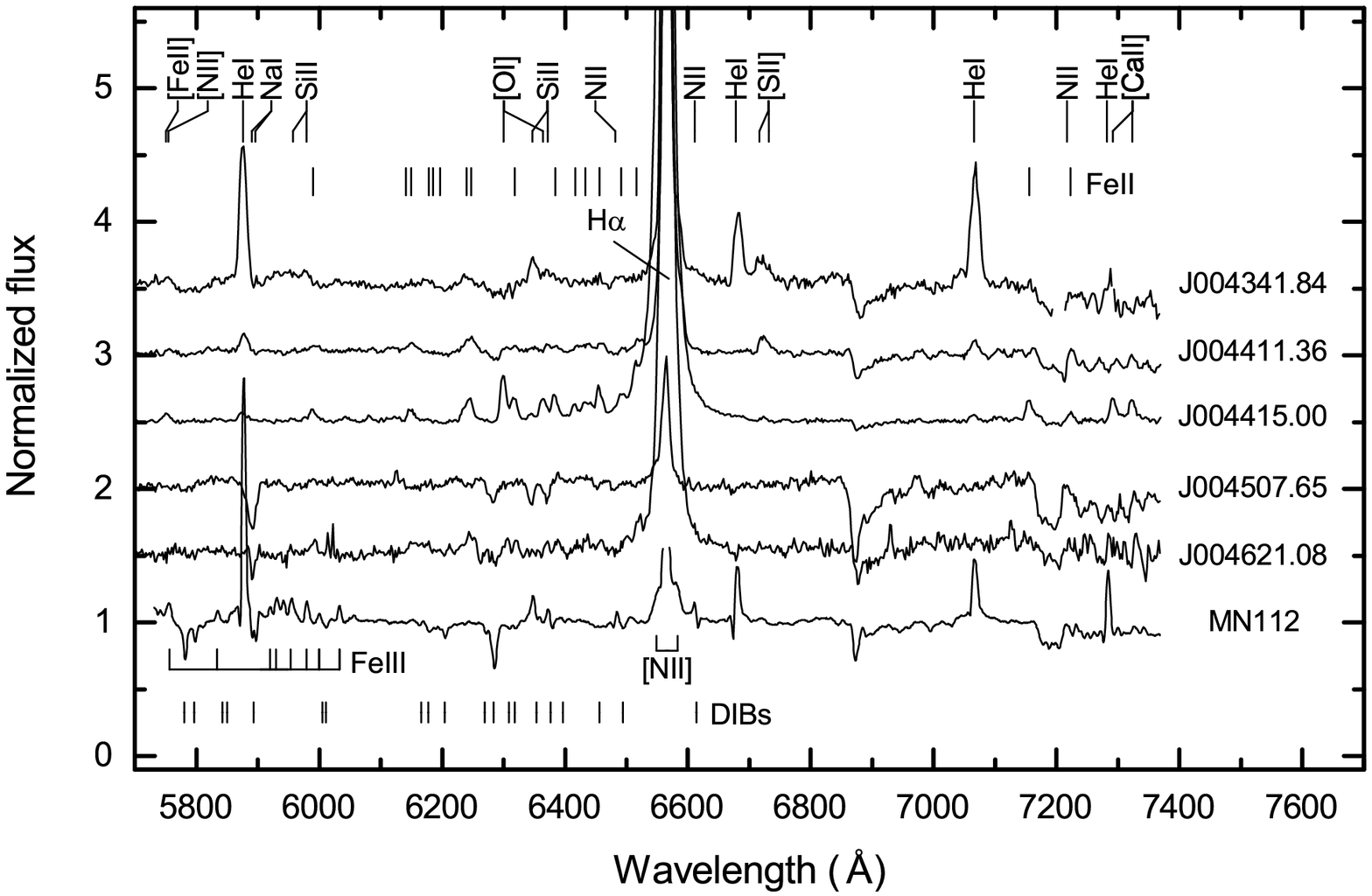}
	\caption{The optical spectra of the J004341.84, J004411.36, J004415.00, J004507.65, J004621.08 and MN112. 
		The principal strong lines and diffuse interstellar bands (DIBs) are identified.}\label{fig:specs_op}

\end{figure*}

\begin{figure*}
	\includegraphics[width=\textwidth]{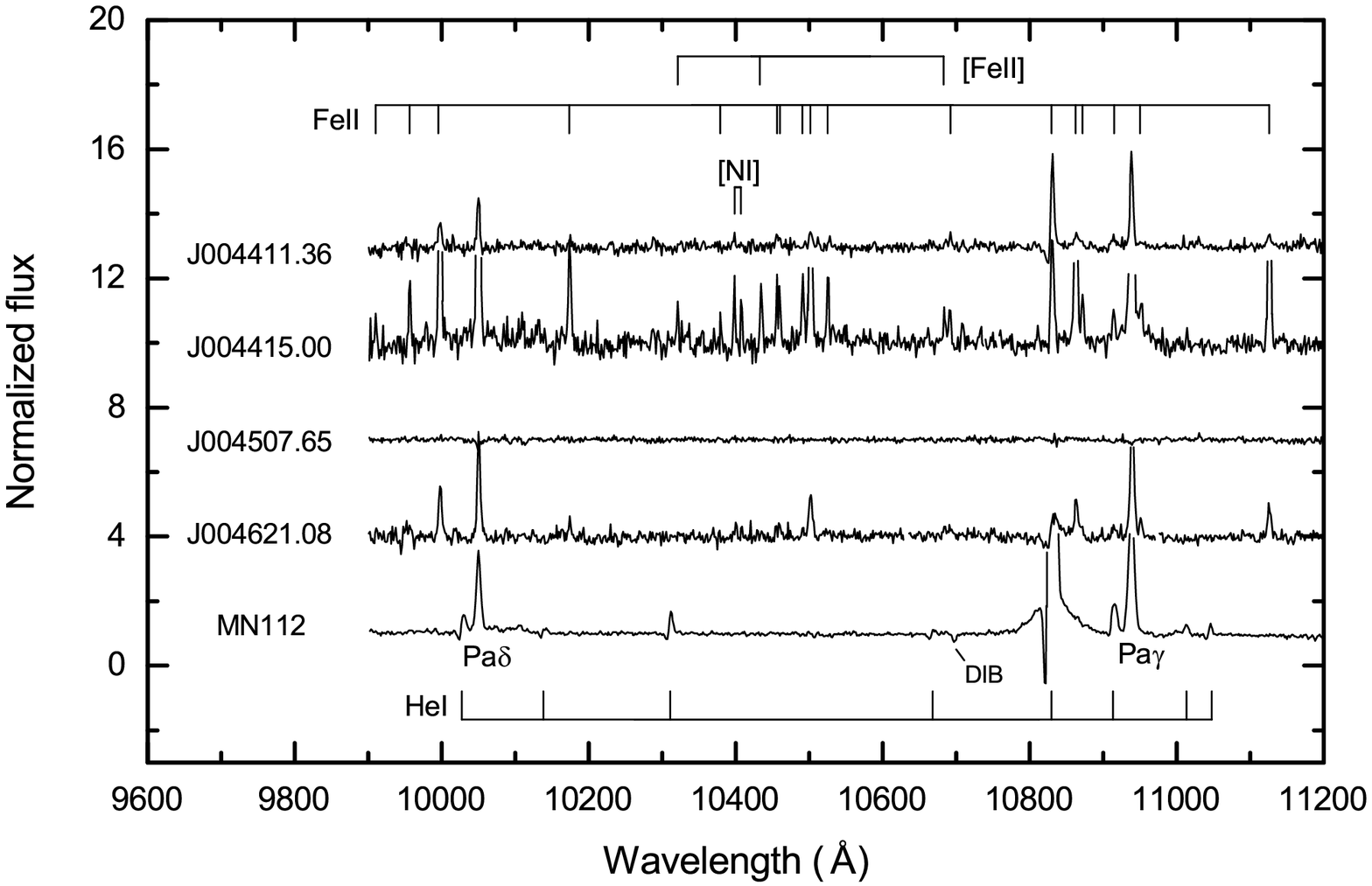}
	\includegraphics[width=\textwidth]{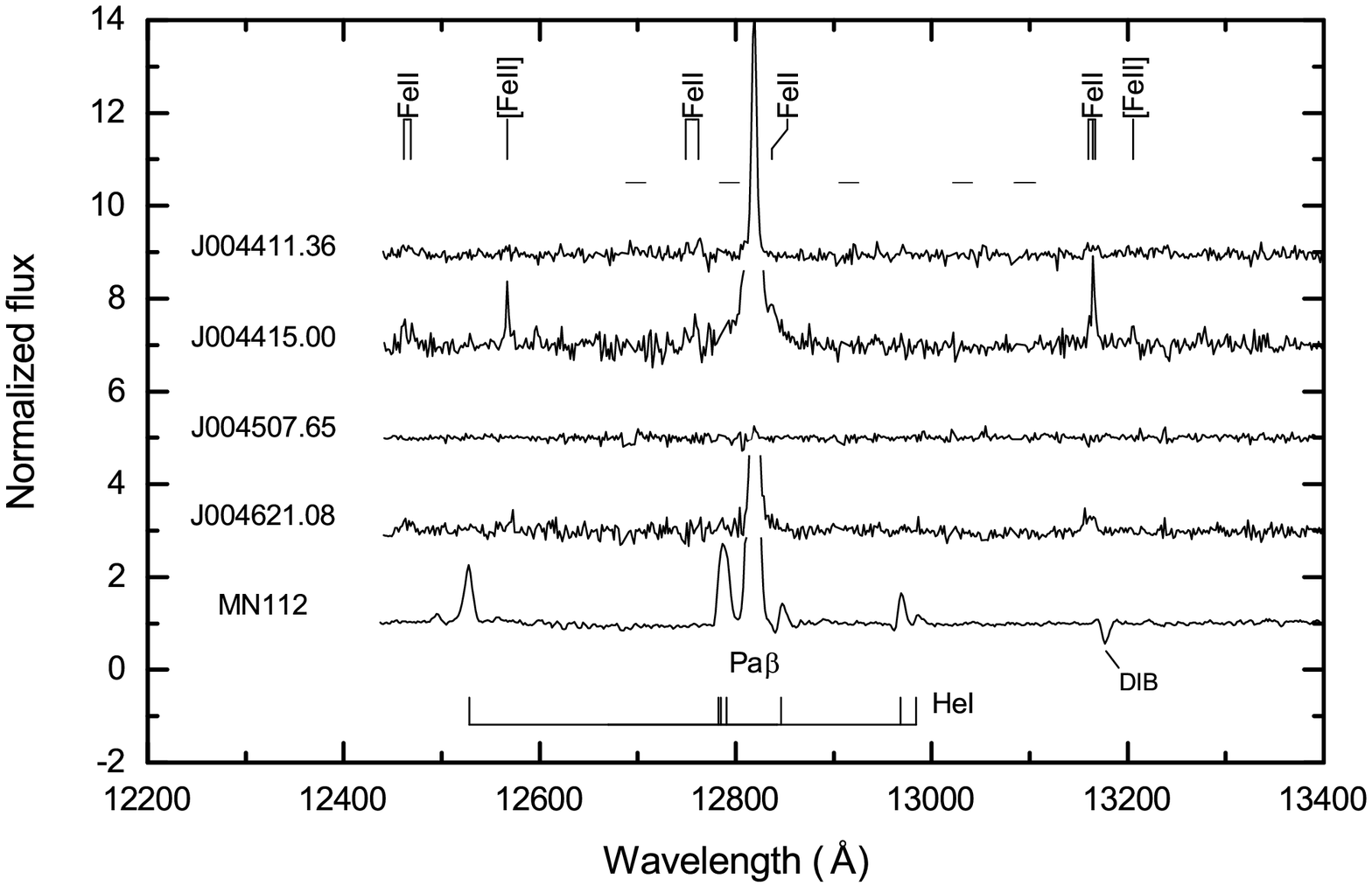}
	\caption{The \textit{J}-band NIR spectra of the same objects as in \autoref{fig:specs_op}.
		The spectrum of J004341.84 is not shown.
		Horizontal unsigned ticks mark the spectra ranges affected by the atmospheric emission and telluric absorption.}\label{fig:specs_ir1}
\end{figure*}

\begin{figure*}
	\includegraphics[width=170mm]{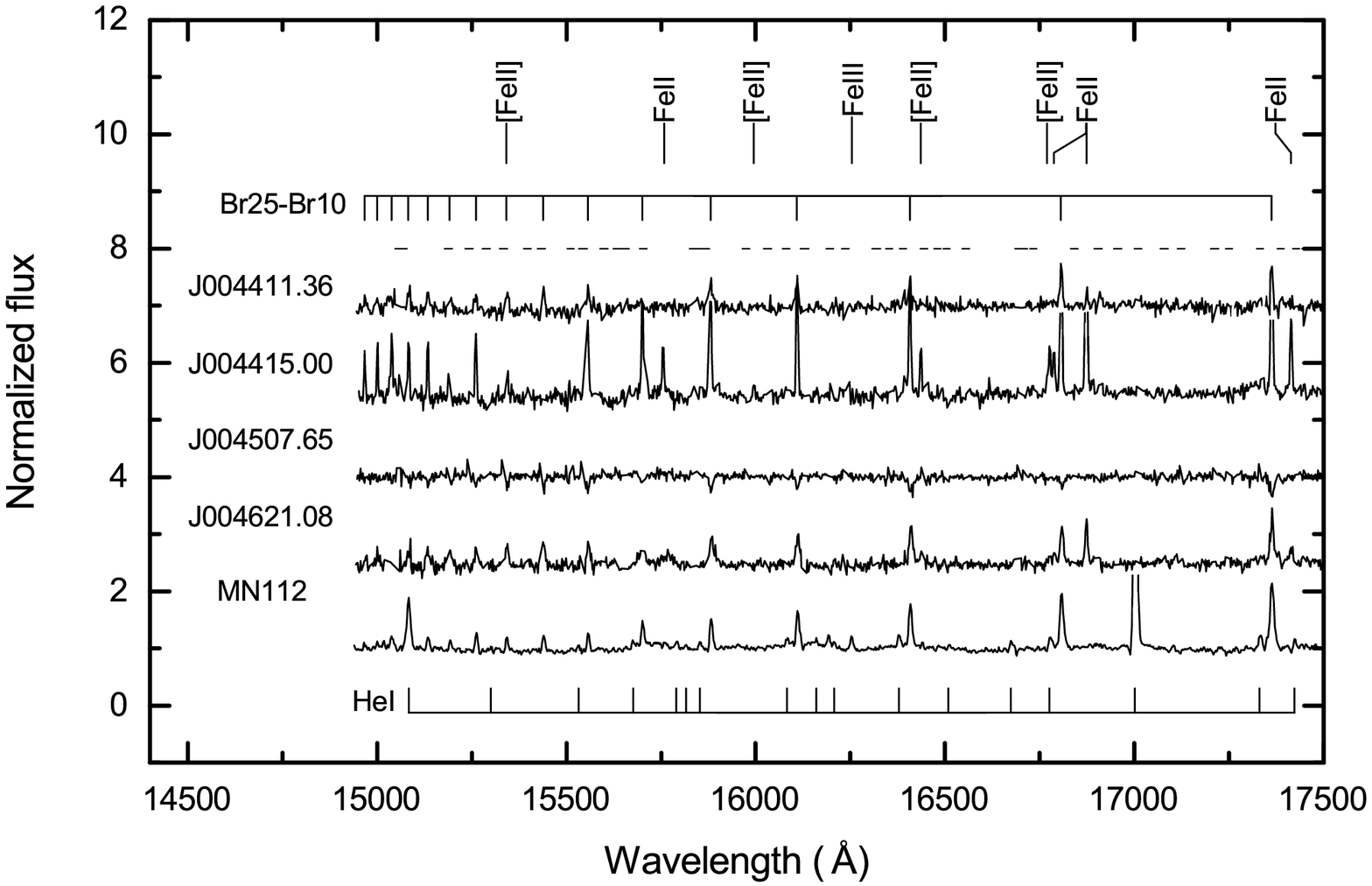}
	\includegraphics[width=170mm]{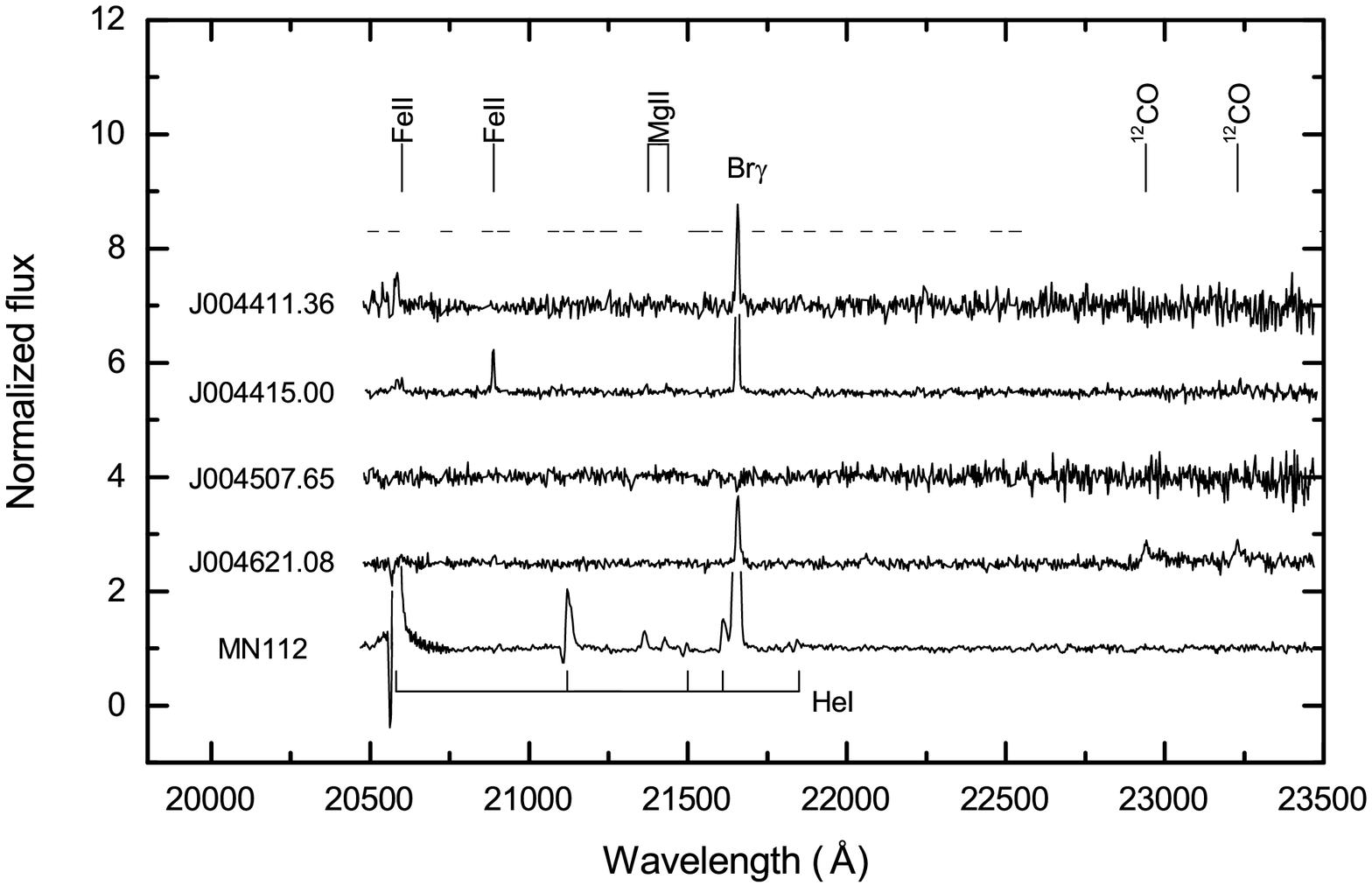}
	\caption{The \textit{H} and \textit{K}-band NIR spectra of the same objects as in \autoref{fig:specs_op}.
		The IR spectrum of J004341.84 is not shown.
	The horizontal unsigned ticks mark the spectra ranges affected by the atmospheric emission and telluric absorption.}\label{fig:specs_ir2}
\end{figure*}

\autoref{fig:specs_op}--\ref{fig:specs_ir2} demonstrates spectra of our stars at six bandpasses. The principal lines are identified and designated: emission lines of Balmer, Paschen and Brackett series, together with \ion{He}{i}, \ion{Fe}{ii} and [\ion{Fe}{ii}], \ion{Fe}{iii}, \ion{Si}{ii}, \ion{N}{ii}, [\ion{N}{ii}] and some other. 

Object J004507.65 shows absorption lines \ion{Si}{ii} (6347, 6371\,\AA). We used MN112 spectrum for comparison since it is almost identical to the spectrum of the well known LBV P Cygni \cite{Massey2006a}. Both the stars J004341.84 and MN112 have P Cygni profiles in hydrogen, \ion{He}{i} and \ion{Fe}{iii} lines, and similarity of their spectra is evident. This supports the LBV classification of J004341.84. J004621.08 has $^{12}$CO lines, which points at the presence of warm stellar wind around the star. We consider the spectra of individual objects in detail below. 

\subsection{The spectral energy distribution}\label{sect:sed}

\begin{figure*}
\includegraphics[angle=270,width=\columnwidth,trim={0 11pt 35pt 24pt},clip]{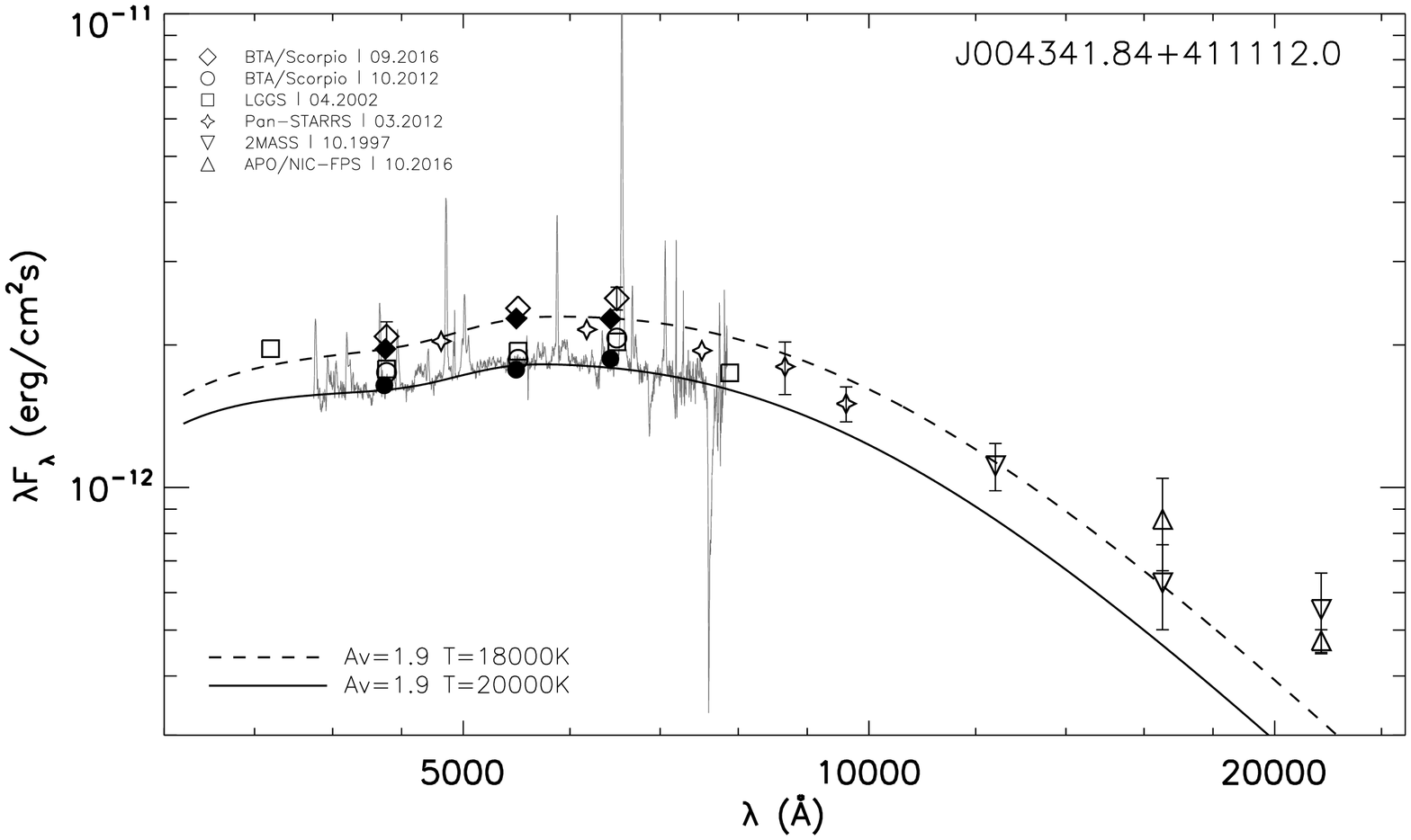}
\includegraphics[angle=270,width=\columnwidth,trim={0 35pt 35pt 0},clip]{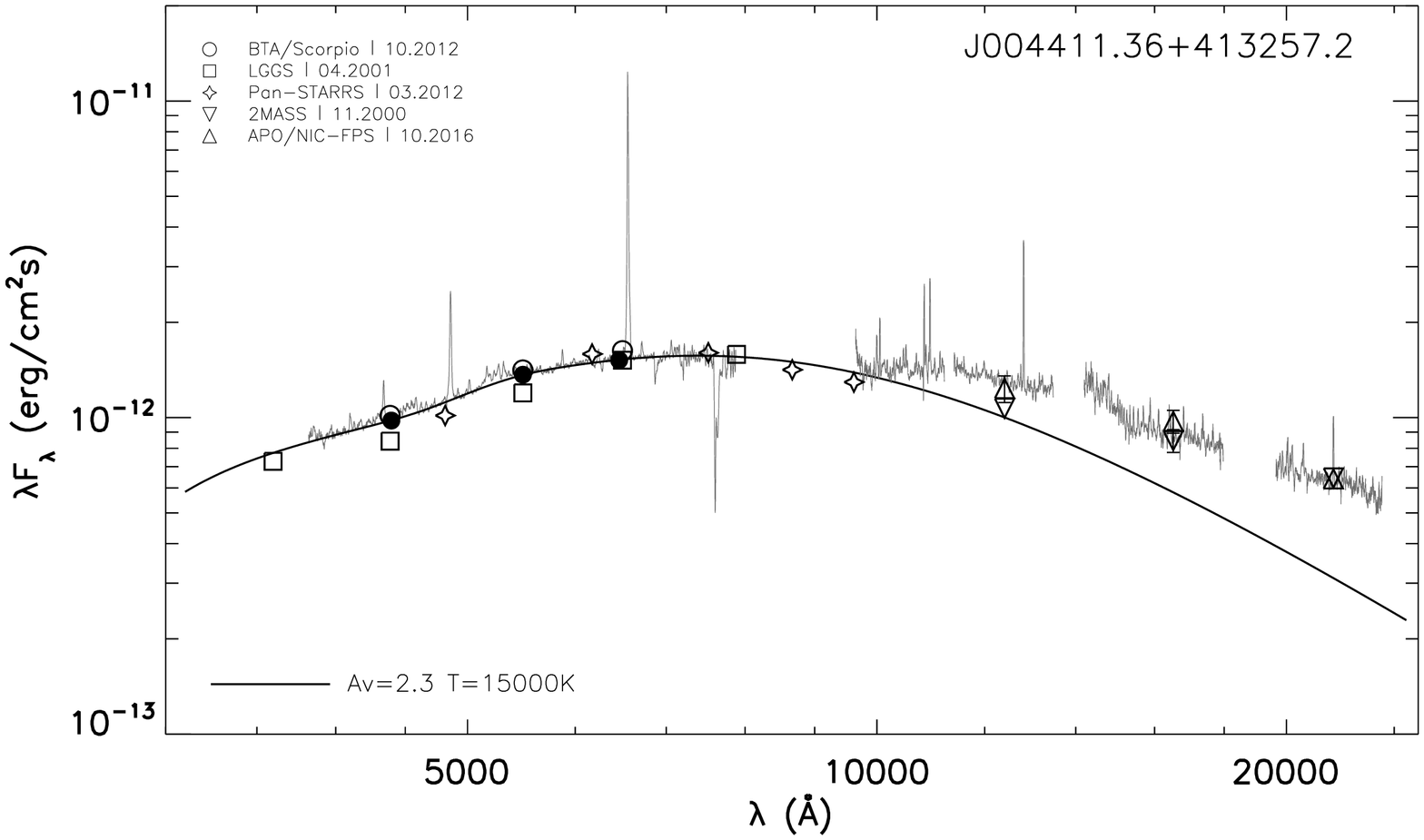}
\includegraphics[angle=270,width=\columnwidth,trim={0 11pt 35pt 24pt},clip]{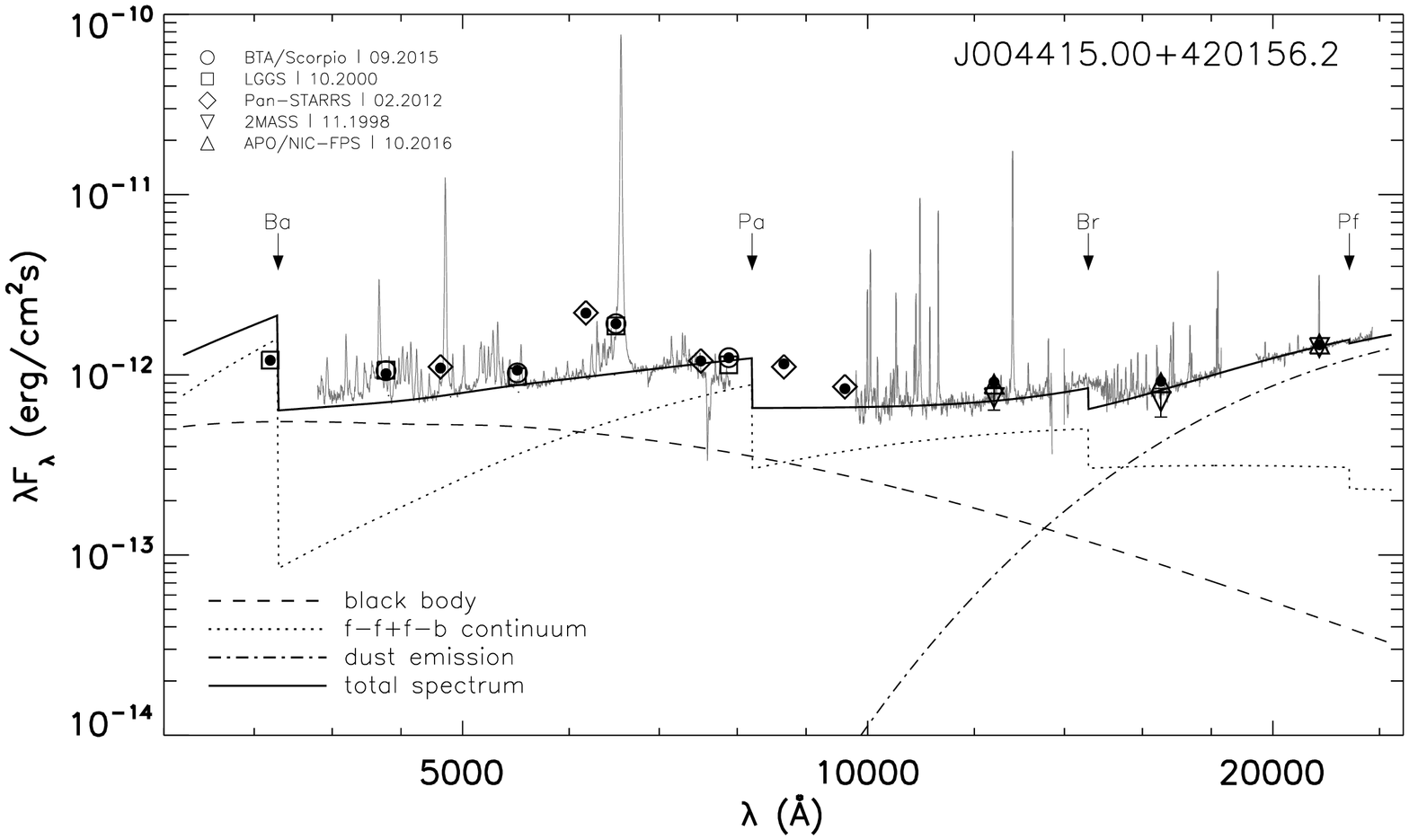}
\includegraphics[angle=270,width=\columnwidth,trim={0 35pt 35pt 0},clip]{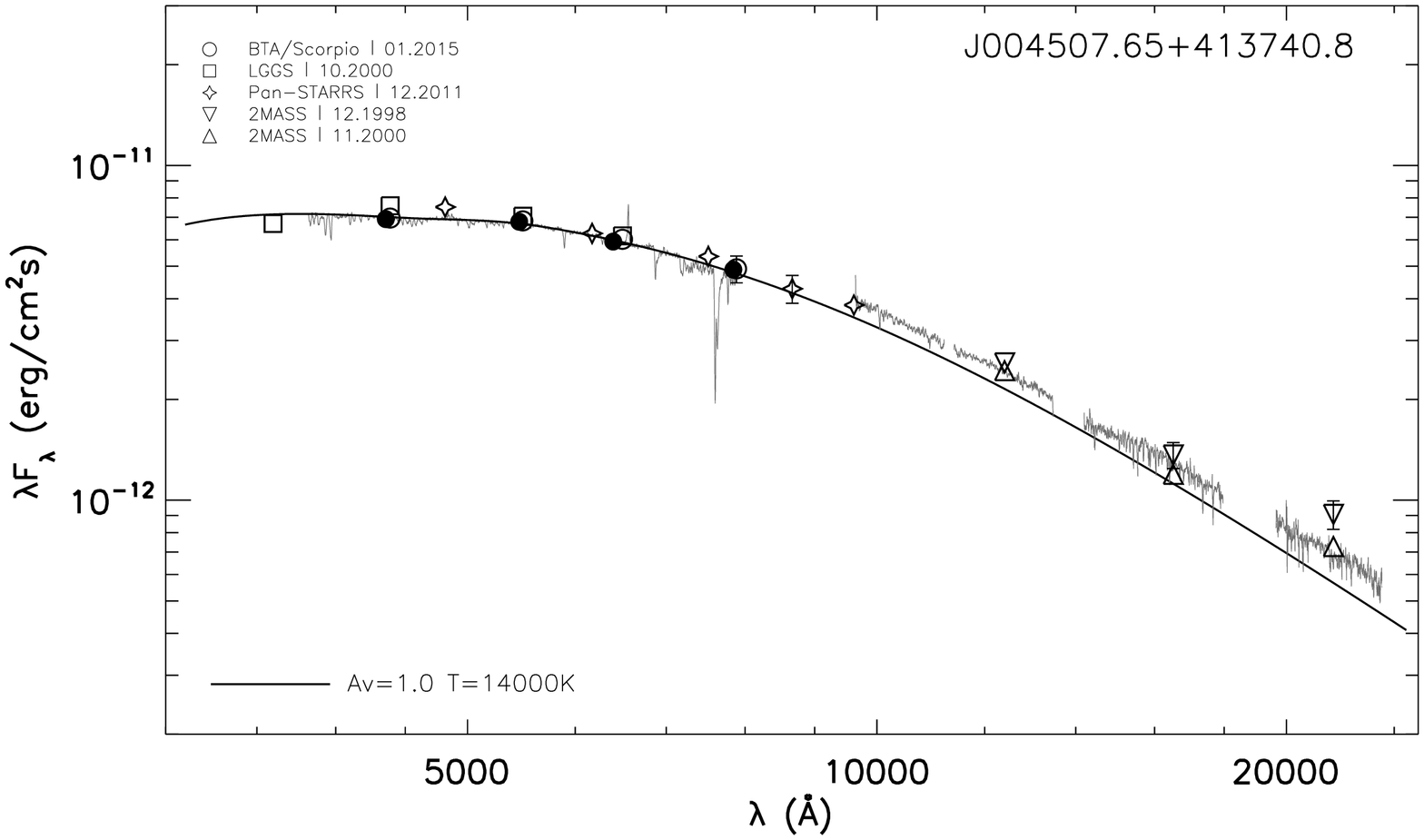}
\includegraphics[angle=270,width=\columnwidth,trim={0 11pt 35pt 24pt},clip]{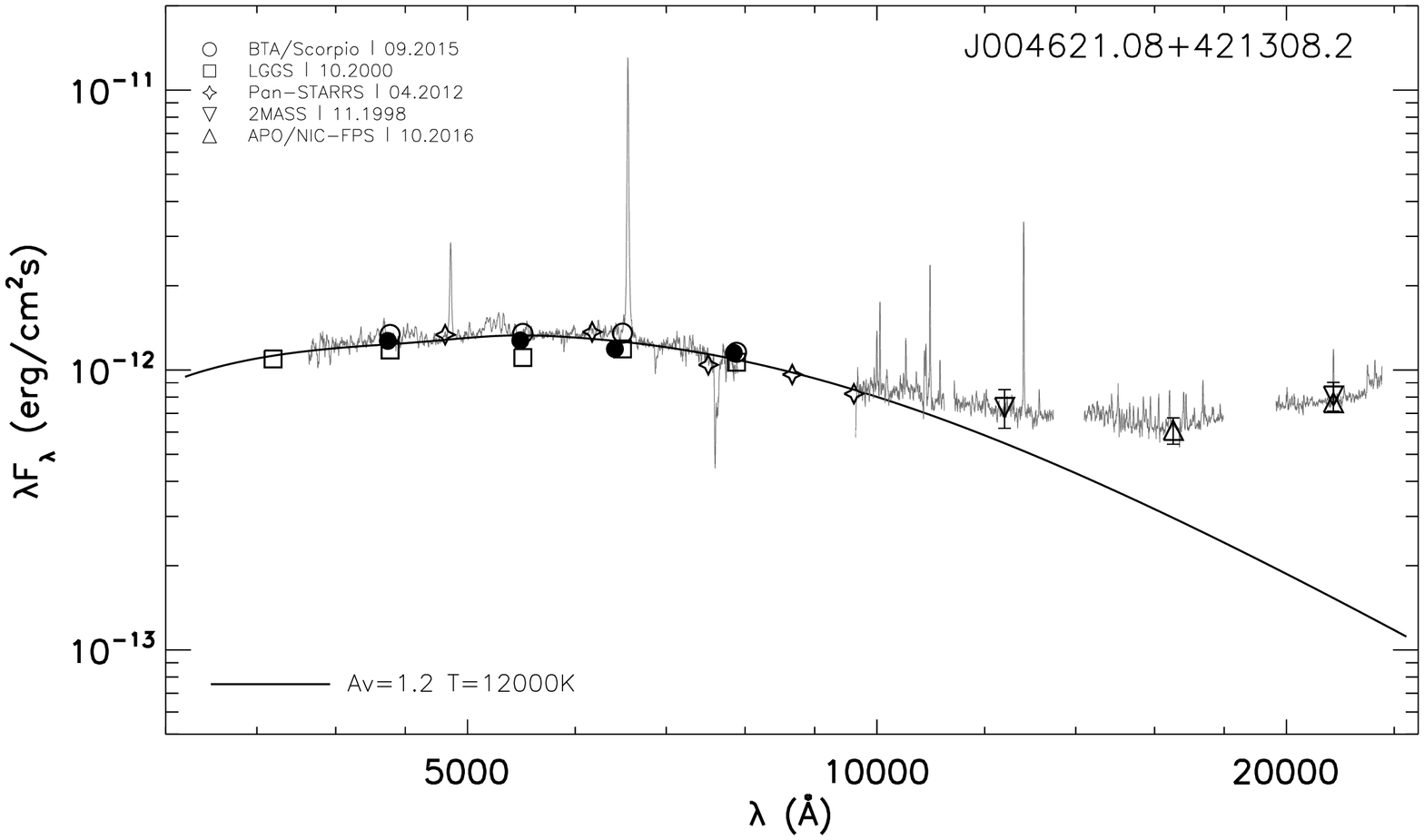}
\includegraphics[angle=270,width=\columnwidth,trim={0 35pt 35pt 0},clip]{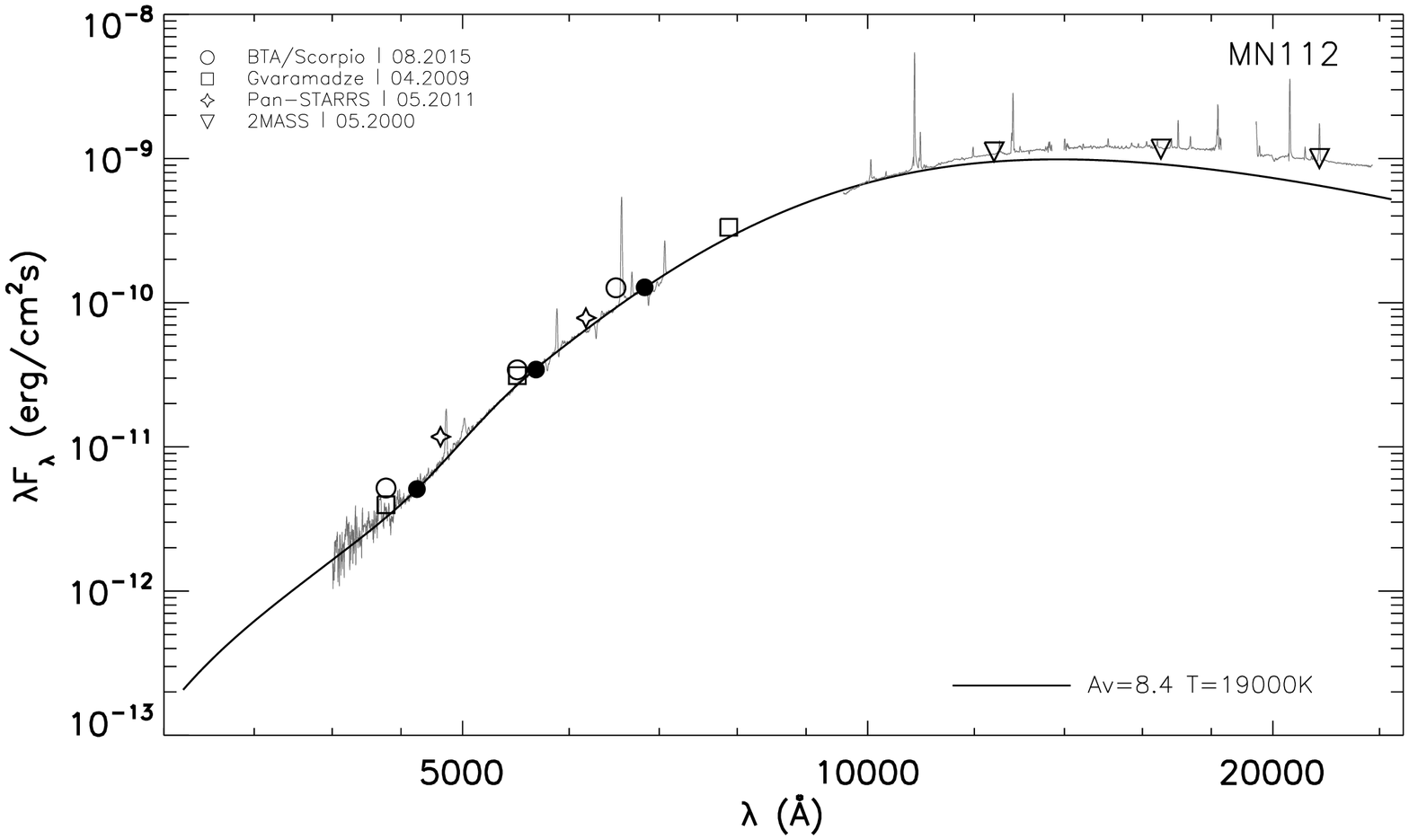}

\caption{The spectra energy distributions of our objects in the optical and NIR ranges. See \autoref{sect:sed} for  details. The photometric points are described in \autoref{tab:photsed}. Note that the error bars for the photometric data points could be less than the symbol size.}\label{fig:seds}
\end{figure*}

\autoref{fig:seds} shows the spectral energy distribution (SED) for our stars in the optical and NIR ranges. The SEDs are made with our spectra and photometry, in a combination with literature data. The used photometric data are summarized in \autoref{tab:photsed}, which shows the date of observations, symbols used in the SED figures, photometric bands, magnitudes with uncertainties and references. The reference source designations are 2MASS -- Two Micron All Sky Survey \protect\citep{Cutri2003}, 2MASS-6X -- Two Micron All Sky Survey 6X \protect\citep{Cutri2012}, LGGS -- Local Group Galaxies Survey \citep{Massey2006}, PS1 -- Panoramic Survey Telescope and Rapid Response part~1 (Pan-STARRS1, \cite{Chambers2016}) and Gv2010 -- \protect\cite{Gvaramadze2010}. The LGGS and PS1 data present the magnitudes averaged over several observational epochs, with averaged observing date. The PS1 data uncertainties are estimated as the r.m.s. of all observed magnitudes. 

\begin{table*}
	\caption{Photometric data for SEDs. The symbols are kept the same as in \autoref{fig:seds}. Also we show the photometric bands, magnitudes of the stars with uncertainties, date of observation, and references to the sources of photometric data. The asterisks at the first column designate the approximated data (see text).}\label{tab:photsed}
	
	\begin{tabular}{lcllc}
		\hline
		\hline 
		Date & Symbol & Filters & Magnitudes & Source \\
		\hline 
		\multicolumn{5}{c}{J004341.84} \\
		\hline
		24.10.1997 & \triangldnsym & \textit{JHK$_s$} & 16.359$\pm$0.124 16.184$\pm$0.221 15.563$\pm$0.210 & 2MASS-6X \\
		06.04.2002 & \squaresym & \textit{UBVRI} & 17.242$\pm$0.007 18.002$\pm$0.006 17.547$\pm$0.004 17.140$\pm$0.006 16.853$\pm$0.007 & LGGS \\	
		25.03.2012 & \starsym & \textit{grizy}$_{\text{PS}}$ & 17.616$\pm$0.020 17.284$\pm$0.023 17.182$\pm$0.036 17.112$\pm$0.138 17.194$\pm$0.092 & PS1 \\
		16.10.2012* & \circlesym & \textit{BVR} & 18.02$\pm$0.05 17.59$\pm$0.03 17.12$\pm$0.05 & this paper \\  
		26.09.2016* & \diamondsym  & \textit{BVR}	& 17.83$\pm$0.08 17.32$\pm$0.05 16.91$\pm$0.06 & this paper \\
		26.10.2016 & \trianglupsym & \textit{HK$_s$} & 15.85$\pm$0.24 15.73$\pm$0.06 & this paper \\
		\hline 
		\multicolumn{5}{c}{J004411.36} \\
		\hline
		14.11.2000 & \triangldnsym & \textit{JHK$_s$} & 16.400$\pm$0.053 15.862$\pm$0.090 15.395$\pm$0.078 & 2MASS-6X \\
		31.01.2001 & \squaresym & \textit{UBVRI} & 18.320$\pm$0.006 18.813$\pm$0.004 18.071$\pm$0.003 17.455$\pm$0.004 16.959$\pm$0.005 & LGGS \\	
		24.03.2012 & \starsym & \textit{grizy}$_{\text{PS}}$ & 18.371$\pm$0.013 17.615$\pm$0.044 17.392$\pm$0.049 17.372$\pm$0.036 17.356$\pm$0.047 & PS1 \\
		17.10.2012* & \circlesym & \textit{BVR} & 18.61$\pm$0.07 17.89$\pm$0.05 17.38$\pm$0.07 & this paper \\
		26.10.2016 & \trianglupsym & \textit{JHK$_s$} & 16.24$\pm$0.11 15.71$\pm$0.09 15.40$\pm$0.06 & this paper \\
		\hline 
		\multicolumn{5}{c}{J004415.00} \\
		\hline
		05.11.1998 & \triangldnsym & \textit{JHK$_s$} & 16.791$\pm$0.160 16.050$\pm$0.196 14.540$\pm$0.076 & 2MASS \\	
		06.10.2000* & \squaresym & \textit{UBVRI} & 17.770$\pm$0.009 18.563$\pm$0.008 18.291$\pm$0.005 17.230$\pm$0.007 17.319$\pm$0.009 & LGGS \\ 
		02.02.2012* & \starsym & \textit{grizy}$_{\text{PS}}$ & 18.276$\pm$0.018 17.257$\pm$0.012 17.710$\pm$0.008 17.638$\pm$0.037 17.801$\pm$0.039 & PS1 \\
		04.09.2015* & \circlesym & \textit{BVRI} & 18.57$\pm$0.05 18.24$\pm$0.02 17.20$\pm$0.04 17.22$\pm$0.03 & this paper \\
		26.10.2016* & \trianglupsym & \textit{JHK$_s$} & 16.63$\pm$0.05 15.86$\pm$0.10 14.52$\pm$0.03 & this paper \\
		\hline 
		\multicolumn{5}{c}{J004507.65} \\
		\hline
		04.12.1998 & \triangldnsym & \textit{JHK$_s$} & 15.448$\pm$0.049 15.343$\pm$0.097 15.025$\pm$0.106 & 2MASS \\
		06.10.2000 & \squaresym & \textit{UBVR} & 15.908$\pm$0.007 16.428$\pm$0.006 16.145$\pm$0.004 15.934$\pm$0.006 & LGGS \\ 
		14.11.2000 & \trianglupsym & \textit{JHK$_s$} & 15.506$\pm$0.031 15.487$\pm$0.061 15.265$\pm$0.063 & 2MASS-6X \\
		27.12.2011 & \starsym & \textit{grizy}$_{\text{PS}}$ & 16.198$\pm$0.010 16.125$\pm$0.051 16.083$\pm$0.017 16.170$\pm$0.104 16.179$\pm$0.011 & PS1 \\
		17.01.2015* & \circlesym & \textit{BVRI} & 16.52$\pm$0.06 16.18$\pm$0.03 15.96$\pm$0.03 15.73$\pm$0.10 & this paper \\
		\hline 
		\multicolumn{5}{c}{J004621.08} \\
		\hline	
		05.11.1998 & \triangldnsym & \textit{JK$_s$} & 16.808$\pm$0.171 15.151$\pm$0.128 & 2MASS \\
		02.10.2000 & \squaresym & \textit{UBVRI} & 17.876$\pm$0.008 18.449$\pm$0.006 18.155$\pm$0.004 17.721$\pm$0.006 17.389$\pm$0.008 & LGGS \\ 
		25.04.2012 & \starsym & \textit{grizy}$_{\text{PS}}$ & 18.073$\pm$0.017 17.778$\pm$0.020 17.858$\pm$0.046 17.790$\pm$0.058 17.849$\pm$0.077 & PS1 \\
		04.09.2015* & \circlesym & \textit{BVRI} & 18.31$\pm$0.03 17.94$\pm$0.05 17.58$\pm$0.03 17.30$\pm$0.03 & this paper \\
		26.10.2016 & \trianglupsym & \textit{HK$_s$} & 16.22$\pm$0.12 15.21$\pm$0.09 & this paper \\
		\hline 
		\multicolumn{5}{c}{MN112} \\
		\hline	
		04.05.2000 & \triangldnsym & \textit{JHK$_s$} & 8.857$\pm$0.021 8.016$\pm$0.024 7.418$\pm$0.021 & 2MASS \\
		22.04.2009 & \squaresym & \textit{BVI} & 17.13$\pm$0.12 14.53$\pm$0.03 11.15$\pm$0.03 & Gv2010 \\
		30.05.2011 & \starsym & \textit{gr}$_{\text{PS}}$ & 15.712$\pm$0.020 13.381$\pm$0.013 & PS1 \\
		17.08.2015* & \circlesym & \textit{BVR} & 16.84$\pm$0.06 14.43$\pm$0.08 12.65$\pm$0.07 & this paper \\
		\hline		
	\end{tabular}

\end{table*}

We approximate the photometric data points (denoted by the asterisks in \autoref{tab:photsed}) with black body spectrum taking into account the dust extinction \citep{Fitzpatrick1999} using $R_V$ = 3.1. We use the \textit{UBVRI} spectra range only for the approximation because the model does not take into account contribution from the bremsstrahlung radiation and the dust emission (except the case of J004415.00, see bellow).  

Most of our objects have strong emission lines in their spectra, which can contribute up to 10--15\ per cent to the photometric SED points. We subtract this contribution from the photometric points using our spectra. We also apply corrections to the average band wavelengths for the spectra slope. Photometric points obtained with those corrections are marked with the filled symbols in \autoref{fig:seds}. For the comparison with existing literature data, we also show not corrected data points with the same but open symbols in \autoref{fig:seds}. The results of the SED approximation are shown with the solid lines and dashed (in the case of fitting two sets of points) lines in \autoref{fig:seds}. Parameters of corresponding models are given in the figure legends and summarized in \autoref{tab:objpars}. 

To show the SEDs based on our optical spectra, we multiply the continuum normalized spectra by best-fit curves of corresponding photometric points because the flux calibration of optical long-slit spectra are affected by the light losses on the slit. Although the applied method of the NIR spectra flux calibration helps better handle the atmospheric extinction \cite{Vacca2003}, our NIR spectra have rather low signal-to-noise (S/N) ratio, so we scale the NIR spectra to tie them to the NIR photometric data points for all objects except the MN112, for which the NIR spectrum was obtained with high S/N ratio. 
 
The parameters $A_V$ and temperature are degenerated in the spectra modeling, which prohibits to determine both of them in a general case. Breaking the degeneracy is possible with some additional constraints on the model parameters. We start with preliminary estimates of the stellar temperature ($T_{\text{sp}}$ in \autoref{tab:objpars}) using the visibility of emission lines \ion{He}{i}, \ion{He}{ii}~4686\,\AA, and \ion{Fe}{ii} in the spectra. Next, we fit the spectrophotometric data using the constrained temperature. This allows us to estimate the $A_V$ with appropriate accuracy, and therefore to estimate the bolometric luminosity of the stars. 
We follow this way for most of our stars: J004411.36, J004507.65, J004621.08 and MN112. \autoref{tab:objpars} shows the estimated parameters: $A_V$, stellar temperature, stellar radii, and absolute and bolometric magnitudes $M_V$ and $M_{bol}$. We accept the distance to M31 of 752$\pm$27 kpc \citep{Riess2012}, and the distance to the MN112 of $6.93^{+2.74}_{-1.81}$ kpc \citep{Bailer-Jones2018}. Note again that the bolometric magnitudes are calculated from the model spectra and do not take into account contributions from the bremsstrahlung or dust emission contribution. 

In addition, variable objects help us solve problems of the degeneration by fitting several data sets with additional constraints on parameters. Thus, variable LBV stars become cooler and brighter in the optical spectral range with a nearly constant bolometric luminosity \citep{HumphreysDavidson1994}. The variability has to have large enough amplitude ($\Delta V \gtrsim 0.2$\,mag) and be relatively slow (at least a few months, \citet{Sholukhova2011}), which makes the constant bolometric luminosity a good approximation $(\sigma T^4 4\pi R^2 = const)$ and allows us to constrain model parameters as $T_i=T_1\sqrt{R_1/R_i}$, where the $1$ and $i$ corresponds to the parameters at the first and further stages of variability, respectively. One more constraint for the model parameters comes from the constant extinction $A_V = const$ independent of the variability stage. One of our objects, J004341.84, indicates the LBV-type variability, which enables us to fit two sets of photometric SEDs (\autoref{fig:seds}) and estimate stellar parameters in two different stages (\autoref{tab:objpars}). 

The object J004415.00 required our special consideration. We notice large variations of spectra at the hydrogen series limits. We designate the limits of the Balmer, Paschen, Brackett and Pfund series in \autoref{fig:seds} for the J004415.00 with vertical arrows. A jump at the Brackett series limit in the IR spectrum can be well seen. The Balmer and Paschen jumps can also be noticed by photometric brightness changes. These inverse jumps at the hydrogen series limits attest to the noticeable contribution of free-free (f-f) and free-bound (f-b) emissions to the spectrum. This confirms the presence of ionized circumstellar envelope where the f-f and f-b emissions originate, which is typical for the B[e]SGs \citep{Zickgraf1985,Zickgraf1986,Zickgraf1989}. To take the contribution of f-f and f-b radiations into account and estimate corresponding spectra, we used \textsc{chianti} package \citep{Dere1997,Landi2013}. We consider the case of isothermal pure hydrogen plasma at the temperature of T$_e$=10000\,K \citep{Lamers1998}. From the spectrum of the star we assume that its effective temperature is T$_{\text{star}}$ =15000\,K and well match the model spectrum to the observations with the following parameters: $T_e=10000$\,K, $EM=1.37\times10^{39}$\,cm$^{-5}$ (emission measure), $T_{\text{star}}=15000$\,K, $R_{\text{star}}=43 R_{\sun}$, $A_V=1.1$. In order to describe the IR excess of the SED we also add hot dust emission to the model spectrum by assuming it to be effective black-body radiation with the temperature of about 1000\,K (the best fit obtained with T$_{\text{dust}}=1050$\,K), which is typical for B[e]SGs hot circumstellar dust \citep{Lamers1998}. The result of the SED fitting is shown in \autoref{fig:seds}: the dashed line shows the blackbody radiation, the dotted line designates the contribution of free-free and free-bound emissions, the dash-dotted line demonstrates the dust emission contribution, the solid line indicates the total model spectrum. All showed spectra include the interstellar extinction (\cite{Fitzpatrick1999}, assuming $R_V$ = 3.1) with $A_V$=1.1 obtained as the best fit parameter. Since the object J004415.00 does not show considerable photometric variability, we incorporate some more published photometric data points for fitting (see \autoref{tab:photsed}) in addition to the BTA data. For better illustration of the matching between the model and observed spectra, the synthetic photometry of the total model continuum multiplied by the normalized spectrum is shown with the filled circles. Note that since the parameters are obtained in the isothermal plasma approximation with fixed temperatures of the star and circumstellar plasma, we report the other parameters derived from them as estimates and do not assess their uncertainties. Nevertheless, our estimates of the model parameters are good enough and allow us to constrain the dust extinction and the luminosity of the object.          
 
\begin{table*}
	\begin{center}
		\caption{Results of our SED fitting to our observations. 
			The columns designate the object name, epoch, stellar photosphere temperature, 
			temperature and extinction $A_{V}$ estimated from the SED fitting, 
			stellar radius in the solar units, $M_V$ and $M_{\text{bol}}$. \label{tab:objpars}}
		\begin{tabular}{cccccccc}
			\hline
			Object & Epoch & $T_{\text{sp}}$ (K) & $T_{\text{SED}}$ (K) & $A_{V}$ (mag) & $R$ ($R_{\sun}$) &$M_V$ (mag) &$M_{\text{bol}}$ (mag) \\
			\hline
			 		           & 16.10.2012 & 18000--22000 & 20000 & 			 					&  87$\pm$6  & -8.7$\pm$0.15 & 					 				 \\[-1ex] 
\raisebox{1.5ex}{J004341.84}   & 26.09.2016 &     --         & 18000 & \raisebox{1.5ex}{1.9$\pm$0.12} & 107$\pm$7  & -9.0$\pm$0.15 &  \raisebox{1.5ex}{-10.3$\pm$0.17} \\    
			J004411.36         & 17.10.2012	& 12000--18000 & 15000 & 2.3$\pm$0.20 					& 120$\pm$12 & -8.8$\pm$0.22 &  -9.8$\pm$0.36 	                 \\ 
			J004415.00         & 04.09.2015 & 13000--17000 & 15000 & 1.1          					& 43         & -7.3          &  -7.5                             \\ 
			J004507.65         & 17.01.2015 & 12000--16000 & 14000 & 1.0$\pm$0.12 					& 158$\pm$10 & -9.2$\pm$0.14 & -10.1$\pm$0.34                    \\ 
			J004621.08         & 04.09.2015 & 10000--15000 & 12000 & 1.2$\pm$0.19 					& 92$\pm$8   & -7.6$\pm$0.21 &  -8.2$\pm$0.41                    \\ 
			MN112              & 17.08.2015 & 18000--22000 & 19000 & 8.4$\pm$0.19 					& 60$\pm$24  & -8.2$\pm$0.88 &  -9.3$\pm$0.91                    \\ 
			\hline
		\end{tabular}
	\end{center}
\end{table*}

\subsection{Individual Objects}

\subsubsection*{J004341.84} 

\cite{Massey2007} and \cite{Humphreys2014} classify this LBV candidate as an Of/late-WN star. \cite{Humphreys2014} found that $T=17300$\,K and $A_V = 0.9$\,mag, and also notice the presence of the bremsstrahlung and dust radiation. \cite{Massey2006a} pointed that the object is an analog of the well known LBV star P Cygni. We also notice wide emission lines H$\alpha$ and H$\beta$, P Cygni profiles in the \ion{Fe}{iii} and \ion{He}{i} lines and close matching of the spectrum to those of P Cygni and P Cygni-like LBV candidate MN112.    

Our SED fitting of two stellar states with constant extinction and bolometric luminosity yields $A_V=1.9$\,mag, $T=20000$\,K for the 16.10.2012 epoch, and $T=18000$\,K for the 26.09.2016 epoch. J004341.84+411112.0 demonstrates photometric variability about 0.27\,mag in \textit{V} band. Despite the brightness variation is not outstanding, it is quite enough to classify the candidate as a LBV in quiescent phase. We do not exclude that J004341.84 is an LBV from its luminosity and variations, although further studies of this object are required. 

\subsubsection*{J004411.36} 

The star is noticed as a variable by \cite{Bonanos2003}, with photometric variability amplitude of 0.15\,mag. 
We also observe a spectral variability. Thus, we identify the \ion{He}{i} 5876\,\AA\ line in our spectra, while it was not seen in the spectrum by \cite{King1998} from 1995. Bright lines [\ion{Ca}{ii}] 7291, 7323\,\AA\ point at a warm dust shell around the star. This object has no infrared rise in \textit{JHK} bands, but only infrared excess.  
\cite{Humphreys2013} classify the star as the Fe\,II emission line type. Although B[e]SGs do not show spectra variability, the J004411.36 evidences spectra variability on the 17 years time range. The spectra variability in B[e]SGs was reported only for S18 by \cite{Clark2013}, with a general doubt of the S18 belonging to the B[e]SGs. 
We classify the J004411.36 as Fe\,II emission star, with a possibility of a dormant LBV. The best SED fitting parameters are $A_V=2.3$\,mag and $T=15000$\,K. 

\subsubsection*{J004415.04}

\cite{Massey2007} classified this star as a hot LBV candidate. It was referred to one of the four confirmed B[e]SGs in M31 by \citet{Kraus2019} in the latest review of B[e]SGs. Our spectrum is almost identical to that from \cite{Massey2007} and is very close to the spectrum by \cite{Humphreys2013}. We notice the presence of strong Balmer emission lines, emission lines \ion{Fe}{ii} and [\ion{Fe}{ii}], bright emission of [\ion{Ca}{ii}] 7291, 7323\,\AA, as well as the absence of \ion{He}{i} lines. We see strong excess in NIR spectrum of the object due to hot circumstellar dust (see \autoref{sect:sed}). So the star satisfies all criteria of the B[e] phenomena \citep{Lamers1998}. Since our estimate of the luminosity of the object is $\log (L/L_{\sun}) = 4.9$\ \citep{Lamers1998}, we confirm its classification as a B[e]SG \citep{Humphreys2017a,Kraus2019}. 

\subsubsection*{J004507.65}  

\cite{Massey2007} suggest that this is a cool LBV star and notice its similarity to Var\,B in M33. Our spectra show the H$\alpha$ emission, H$\beta$ filled in by emission and H$\gamma$ in absorption. The \ion{Fe}{ii} and \ion{Si}{ii} 6350\,\AA\ lines are also in the absorption. We do not see any \ion{He}{i} lines. In the IR spectrum we have no excess and as well no emission lines [\ion{Ca}{ii}] and \ion{Ca}{ii}. In the optical spectrum J004507.65 is similar to the V532 (Romano's star) at the maximum in 1992 \citep{Szciefert1996,Sholukhova2011}. However, the variability in the optical range between 2000 and 2017 in archival data and in our data is not detected. The best SED fitting model has parameters $A_V$=1.0\,mag and T=14000\,K. \cite{Humphreys2013} classify this star as an intermediate supergiant A5--A8 with the H$\alpha$ emission. We suggest that it is a warm hypergiant because it shows a high bolometric luminosity $M_{bol} \approx -10.2$ and a large radius $R/R_{\odot} = 158 \pm 10$.

\subsubsection*{J004621.08} 

This star was determined as a LBV candidate by \cite{King1998} and \cite{Massey2007}. We do not detect any variability in our data. \cite{Gordon2016} concluded that the spectrum of this star corresponds to the late A-type supergiant with strong hydrogen emission lines with broad wings. The red part of its spectrum has \ion{Ca}{ii}, [\ion{Ca}{ii}] and $^{12}$CO lines, which indicates of a low density nebula around. Our spectra show broad H$\alpha$ and H$\beta$, and a large number of \ion{Fe}{ii} and [\ion{Fe}{ii}] emission lines. In the \textit{JHK} bands we see an infrared excess, but the spectrum is flattened. The star is classified as a warm hypergiant, which is consistent with the classification by \cite{Humphreys2017a}. The best fit of the SED obtained with $A_V=1.2$\,mag and $T=12000$\,K.

\subsubsection*{MN112} 

\cite{Gvaramadze2010} found an IR nebula near the MN112 and its spectral similarity to P\,Cyg, and classified the star as a LBV candidate. The optical spectrum of MN112 has broad H$\alpha$ and H$\beta$ lines, very bright \ion{He}{i}, and bright \ion{Fe}{iii}, \ion{N}{ii}, \ion{Si}{ii} and \ion{C}{ii} lines. A large number of diffuse interstellar bands is also seen in the spectra. We obtain the first NIR spectrum of MN112 which also shows strong Paschen and Brackett lines, \ion{He}{i} lines with P\,Cygni profiles and \ion{Mg}{ii} lines in \textit{K}-band. 
The distance to the MN112 is $6.93^{+2.74}_{-1.81}$\,kpc, according to GAIA DR2 data \citep{Bailer-Jones2018}, 
which helps us determine the extinction and temperature as $A_V=8.4$\,mag and $T=19000$\,K, respectively.  
Despite the non detection of the brightness variability, we still consider this object as a LBV candidate and look forward to obtain more photometric data to verify its classification. 

\section{Conclusions}

We undertook simultaneous O/NIR spectral and photometric observations of the most massive stars in the Local galaxy M31 and one star in Milky Way, which could be LBV stars, B[e]-supergiants or warm hypergiants. For all the stars we also incorporated archival photometric and spectral data. We estimated the stellar parameters: spectral energy distribution, temperature, reddening, luminosities and radii in order to perform the classification of the stars.  

We continue to apply new method of LBV SED modeling \citep{Sholukhova2015} to studying new LBV candidates. The method assumes applying a constant bolometric luminosity and dust extinction with variable visual brightness. Under these assumptions we break the $A_V$ -- temperature parameters degeneracy by considering the stars in two or more stages. 

All our LBV candidates show optical and NIR spectra typical for LBVs, warm hypergiants or B[e]SGs. Their luminosity corresponds to known LBVs in M31 \citep{Humphreys2014}. We confirm that the star J004415.04 is a B[e]SG.
The objects J004411.36 is classified as a Fe\,II star. We identified two warm hypergiants, J004621.08 and J004507.65.

The star J004341.84 is variable according to our and literature data, and can be classified as a LBV star. For the MN112 where we have got identical spectra with the same temperature, although the J004341.84 is more luminous than MN112.

\section*{Acknowledgements}

This work was supported by the Russian Foundation for Basic Research N 19-02-00432, N 19-52-18007. This research was supported by the Russian Science Foundation (grant N 14-50-00043) in the data reduction. Based on observations obtained with the Apache Point Observatory 3.5-meter telescope, which is owned and operated by the Astrophysical Research Consortium. S.F. and P.N. acknowledge partial support from DN18/10 from the NSF of Bulgaria. O.Sh. and P.N. acknowledge partial support from Ministry of Education and Science of the Republic of Bulgaria, National RI Roadmap Project DO1-157/28.08.2018. The spectrum of J004415.00 was taken within the search for QSOs behind the disk of M31 galaxy supported by short-term scholar Fulbright grant of one of us (P.N.) at the Department of Astronomy, University of Washington for the academic year 2018-2019 by the US Department of States program number G-1-00005. The authors also thank the anonymous referee for their careful review and helpful suggestions that improved the manuscript.

\section*{Data availability}

The data underlying this article will be shared on reasonable request to the corresponding author.

%%%%%%%%%%%%%%%%%%%%%%%%%%%%%%%%%%%%%%%%%%%%%%%%%%



\bibliographystyle{mnras}
\bibliography{LBV_candidates_in_M31} % if your bibtex file is called example.bib


%\begin{thebibliography}{99}
%\end{thebibliography}

%%%%%%%%%%%%%%%%%%%%%%%%%%%%%%%%%%%%%%%%%%%%%%%%%%

%%%%%%%%%%%%%%%%% APPENDICES %%%%%%%%%%%%%%%%%%%%%

%\appendix

%%%%%%%%%%%%%%%%%%%%%%%%%%%%%%%%%%%%%%%%%%%%%%%%%%

% Don't change these lines
\bsp	% typesetting comment
\label{lastpage}
\end{document}